\documentclass[twocolumn,showpacs,preprintnumbers,amsmath,amssymb]{revtex4}
\usepackage{graphicx}
\usepackage{amssymb,latexsym}
\begin{document}
   \title{Comment on ``$w$ and $w'$ of scalar field models of dark energy''}
   \author{Chien-Wen Chen$^{1}$}
     \email{f90222025@ntu.edu.tw}

   \author{Je-An Gu$^{1}$}

   \author{Pisin Chen$^{1,2,3,4}$}

     \affiliation{%
1. Leung Center for Cosmology and Particle Astrophysics, National Taiwan University, Taipei, Taiwan 10617\\
2. Department of Physics, National Taiwan University, Taipei, Taiwan 10617\\
3. Graduate Institute of Astrophysics, National Taiwan University, Taipei, Taiwan 10617\\
4. Kavli Institute for Particle Astrophysics and Cosmology, SLAC
National Accelerator Laboratory, Menlo Park, CA 94025, U.S.A.
}%

   \date{\today}

\begin{abstract}
We comment on the calculation mistake in the paper ``$w$ and $w'$ of
scalar field models of dark energy'' by Takeshi Chiba, where $w$ is
the dark energy equation of state and $w'$ is the time derivative of
$w$ in units of the Hubble time. The author made a mistake while
rewriting the phantom equation of motion, which led to an incorrect
generic bound for the phantom model and an incorrect bound for the
tracker phantom model on the $w\textrm{--}w'$ plane.
\end{abstract}

 \maketitle

In the paper ``$w$ and $w'$ of scalar field models of dark
energy''~\cite{Chiba:2005tj}, Takeshi Chiba derived bounds on $w'$
as a function of $w$ for scalar field models of dark energy
including quintessence, phantom and k-essence. This provides very
useful means to discriminate between dark energy models and to
confront dark energy models with the observational results
(see~\cite{Chen:2009bca} for example). In~\cite{Chiba:2005tj}, we
find that there is a mistake in calculation for the phantom case,
which leads to the incorrect bounds on $w'$. Following the framework
in~\cite{Chiba:2005tj}, we derive the corrected bounds for the
phantom model. In addition, we also correct an argument leading to
the bound for the tracker quintessence.

\section{LIMITS OF PHANTOM}

\subsection{Generic bound}
The phantom model is scalar field dark energy having negative
kinetic term and its equation of state $w<-1$. The energy density
and the pressure are given by $\rho=-\dot\phi^2/2+V$ and
$p=-\dot\phi^2/2-V$, respectively. The equation of motion is given
by
\begin{equation}\label{q1}
\ddot\phi+3H\dot\phi-V_{,\phi}=0 \ .
\end{equation}
Therefore, the phantom field tends to roll up the potential $V$. The
phantom equation of motion can be rewritten as~\cite{Kujat:2006vj}
\begin{equation}\label{q2}
\pm {V_{,\phi}\over V}=\sqrt{{-3\kappa^2(1+w)\over \Omega_{\phi}}}
\left(1+{x'\over 6}\right) \ ,
 \end{equation}
where the plus sign corresponds to $\dot\phi>0$ and the minus sign
to the opposite, $\kappa^2=8\pi G$ and $\Omega_{\phi}$ is the
fractional energy density of the phantom field. The variable $x$ is
defined as
\begin{equation}\label{q3}
 x=\ln\left(-{1+w\over 1-w}\right) \ ,
\end{equation}
and $x'$ is the derivative of $x$ with respect to $\ln a$, which is
related with $w'$ as
\begin{equation}\label{q4}
  x'={2w'\over (1-w)(1+w)} \ .
\end{equation}
Since the left-hand side of Eq.~(\ref{q2}) is positive for the
up-rolling phantom field, we have $1+x'/6>0$. Therefore, using
Eq.~(\ref{q4}), the upper bound on $w'$ is obtained as
\begin{equation}\label{q5}
w'<-3(1-w)(1+w).
\end{equation}
Note that the upper bound on $w'$ for phantom is smoothly connected
to the lower bound on $w'$ for quintessence, the latter of which is
shown as Eq.~(2.6) in~\cite{Chiba:2005tj}.

In~\cite{Chiba:2005tj}, there is a calculation mistake that the
rewritten equation of motion is obtained as
\begin{equation}\label{q6}
\mp {V_{,\phi}\over V}=\sqrt{{-3\kappa^2(1+w)\over \Omega_{\phi}}}
\left(-1+{x'\over 6}\right) \ .
\end{equation}
As a consequence, an incorrect bound on $w'$ is obtained as
\begin{equation}\label{q7}
w'>3(1-w)(1+w).
\end{equation}

\subsection{Tracker phantom}\label{TP}
Following the approach in~\cite{Chiba:2005tj}, the bound can be
tightened for the tracker field. Taking the derivative of
Eq.~(\ref{q2}) with respect to $\phi$, we obtain the tracker
equation for the phantom field
\begin{eqnarray}\label{q8}
\Gamma-1  &=&
\frac{3(w_B-w)(1-\Omega_{\phi})}{(1+w)(6+x')}-\frac{(1-w)x'} {
2(1+w)(6+x')} \nonumber \\  & \ & -\frac{2x''}{(1+w)(6+x')^2} \ ,
\end{eqnarray}
where $\Gamma=VV_{,\phi\phi}/V_{,\phi}^2$, $w_B$ is the equation of
state of the background matter, and $x''$ is the second derivative
of $x$ with respect to $\ln a$. Therefore, for the tracker solution
where $w$ is nearly constant and $\Omega_{\phi}$ is initially
negligible, $w$ is given by
\begin{equation}\label{q9}
 w={w_B-2(\Gamma -1)\over 2(\Gamma -1) +1} \ .
\end{equation}
Thus $\Gamma<1/2$ is required for tracker phantom, which has $w<-1$.

Following~\cite{Chiba:2005tj}, we consider a solution in which
initially $w$ follows the tracker solution in Eq.~(\ref{q9}) and
then evolves toward $-1$. Therefore, the tracker $w$ in
Eq.~(\ref{q9}) is a lower bound of $w$. In such solution, $x'$
eventually stops decreasing and then increases back to a value near
zero. The minimum value of $x'$, $x_m'$, gives an upper bound on
$w'$ via Eq.(\ref{q4}). To find $x_m'$, we put $x''=0$ and $w_B=0$
in Eq.~(\ref{q8}) and find that
\begin{equation}\label{q10}
x_m'=-6{w(1-\Omega_{\phi})+2(1+w)(\Gamma -1)\over
(1-w)+2(1+w)(\Gamma -1)} \ .
\end{equation}
Since $x_m'$ is an increasing function of $w$, a lower bound on
$x_m'$ is given by that of $w$, for which we take the tracker $w$ in
Eq.~(\ref{q9}) and obtain
\begin{equation}\label{q11}
x_m'>\frac{6w\Omega_{\phi}}{1-2w}>\frac{6w}{1-2w} \ .
\end{equation}
With Eq.~(\ref{q4}), we then obtain the upper bound on $w'$
\begin{equation}\label{q12}
w'<\frac{3w(1-w)(1+w)}{1-2w} \ .
\end{equation}

In \cite{Chiba:2005tj}, there is a mistake that the tracker equation
for the phantom field is given as
\begin{eqnarray}\label{q13}
\Gamma-1  &=&
\frac{3(w_B-w)(1-\Omega_{\phi})}{(1+w)(6-x')}-\frac{(1-w)x'} {
2(1+w)(6-x')} \nonumber \\  & \ & +\frac{2x''}{(1+w)(6-x')^2} \ .
\end{eqnarray}
As a consequence, the resulting $x_m'$ is
\begin{eqnarray}\label{q14}
x_m'&=&-6{w(1-\Omega_{\phi})+2(1+w)(\Gamma -1)\over
(1-w)-2(1+w)(\Gamma -1)}   \\&>&6w\Omega_{\phi}>6w.
\end{eqnarray}
Therefore, an incorrect bound on $w'$ is given as
\begin{equation}\label{q15}
w'<3w(1-w)(1+w).
\end{equation}

\begin{figure}
\includegraphics[width=8.3cm]{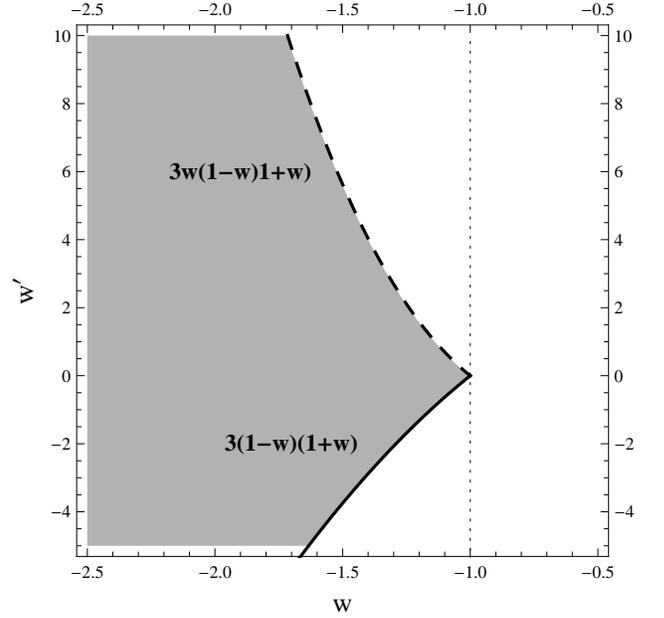}
  \caption{\label{fig1}Bounds on $w'$ as a function of $w$ for the phantom
  model from \cite{Chiba:2005tj}. The solid curve is the generic lower bound
  of the phantom. The dashed curve is the upper bound of the tracker
  phantom. The allowed region for the tracker phantom is filled with
  the grey color.
  }
\end{figure}

\begin{figure}
\includegraphics[width=8.3cm]{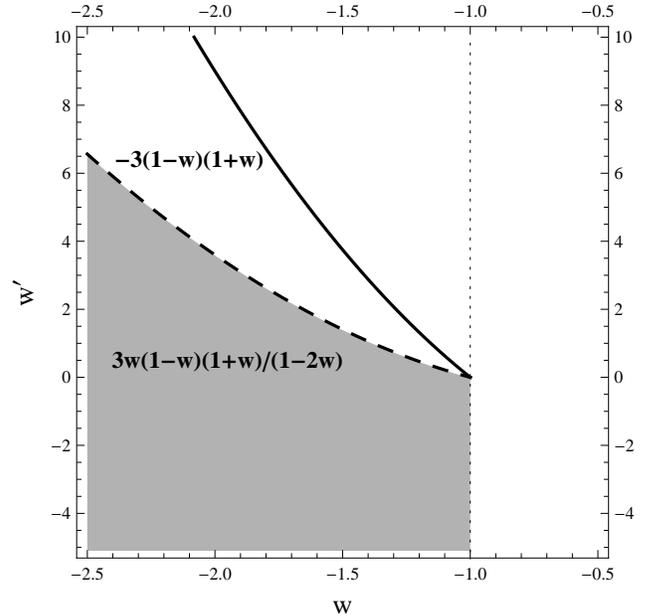}
  \caption{\label{fig2}Corrected bounds we obtain on $w'$ as a function of $w$ for the phantom
  model. The solid curve is the generic upper bound of
  the phantom. The dashed curve is the upper bound of the tracker
  phantom. The allowed region for the tracker phantom is filled with
  the grey color.}
\end{figure}

The bounds for the phantom model from \cite{Chiba:2005tj} are shown
in Fig.~\ref{fig1} and the corrected ones we obtain are shown in
Fig.~\ref{fig2}.

\section{LIMITS OF TRACKER QUINTESSENCE}
For tracker quintessence the Eqs.~(\ref{q8}) \textrm{--} (\ref{q11})
remain valid. However, we have $w>-1$ and $\Gamma> 1$ in this case.
In addition, instead of an upper bound on $w'$, the minimum value of
$x'$, $x'_m$, gives a lower bound on $w'$ via Eq.~(\ref{q4}). Since
$x_m'$ is a decreasing function of $w$, a lower bound on $x_m'$ is
given by an upper bound of $w$, for which we take the tracker $w$ in
Eq.~(\ref{q9}) and obtain Eq.~(\ref{q11}). As a result, the lower
bound on $w'$ is given as
\begin{equation}\label{q16}
w'>\frac{3w(1-w)(1+w)}{1-2w} \ .
\end{equation}

Note that our argument which leads to the lower bound on $x_m'$ is
different from that in~\cite{Chiba:2005tj}, between Eq.~(2.9) and
Eq.~(2.10): ``since $x_m$ is an increasing function of $w(<-1)$, a
lower bound is given by $w$ of the tracker solution Eq.~(2.8)''.
However, the resulting bound is the same.
\section*{ACKNOWLEDGMENT}
C.-W.~Chen is supported by the Taiwan National Science Council (NSC)
under Project No.~NSC 95-2119-M-002-034 and NSC
96-2112-M-002-023-MY3, Gu under NSC 98-2112-M-002-007-MY3, and P.
Chen by NSC under Project No.~NSC 97-2112-M-002-026-MY3 and by US
Department of Energy under Contract No.~DE-AC03-76SF00515. All the
authors thank Leung Center for Cosmology and Particle Astrophysics
for the support.

\end{document}